\documentclass[aps,twocolumn,showpacs,preprintnumbers,amsmath,amssymb,nofootinbib,showkeys]{revtex4-1}

\usepackage{epsfig,graphicx,amsmath,amssymb}
\usepackage{subcaption}
\usepackage{picture}
\usepackage{slashed}
\usepackage{color}
\usepackage{soul}

\newcommand{\dd}{\mathrm{d}}

\begin{document}

\title{Phenomenology of bivariate approximants: the $\pi^0 \to e^+e^-$ case and its impact on the electron and muon $g-2$}

\author{Pere Masjuan} \email{masjuan@kph.uni-mainz.de}
\author{Pablo Sanchez-Puertas} \email{sanchezp@kph.uni-mainz.de}
\affiliation{PRISMA Cluster of Excellence, Institut f\"ur Kernphysik, Johannes Gutenberg-Universit\"at Mainz,
D-55099 Mainz, Germany}
\preprint{MITP/15-028}
\begin{abstract}
The current 3$\sigma$ discrepancy  between experiment and Standard Model predictions for $\pi^0 \to e^+e^-$ is reconsidered using the Pad\'e Theory for bivariate functions, the Canterbury approximants. This method provides a model-independent data-driven approximation to the decay as soon as experimental data for the doubly virtual $\pi^0$ transition form factor are available. It also implements the correct QCD constraints of the form factor both at low- and high-energies. We reassess the Standard Model result including, for the first time, a systematic error. Our result, BR$(\pi^0 \to e^+e^-)=6.23(5)\times 10^{-8}$, still represents a discrepancy larger than $2\sigma$, unsurmountable with our present knowledge of the Standard Model, and would claim New Physics if the experimental result is confirmed by a new measurement. Our method also provides the adequate tool to extract the doubly virtual form factor from experimental data in a straightforward manner. This measurement would further shrink our error and establish once and for all the New Physics nature of the discrepancy. In addition, we remark the challenge this discrepancy poses in the evaluation of the hadronic light-by-light scattering contribution to the $(g-2)_{\mu}$, specially confronted with the foreseen accuracy of the forthcoming $(g-2)_{\mu}$ experiments.

\end{abstract}


\maketitle

\section{Introduction}

\begin{figure}[h]
\includegraphics[width=0.35\textwidth]{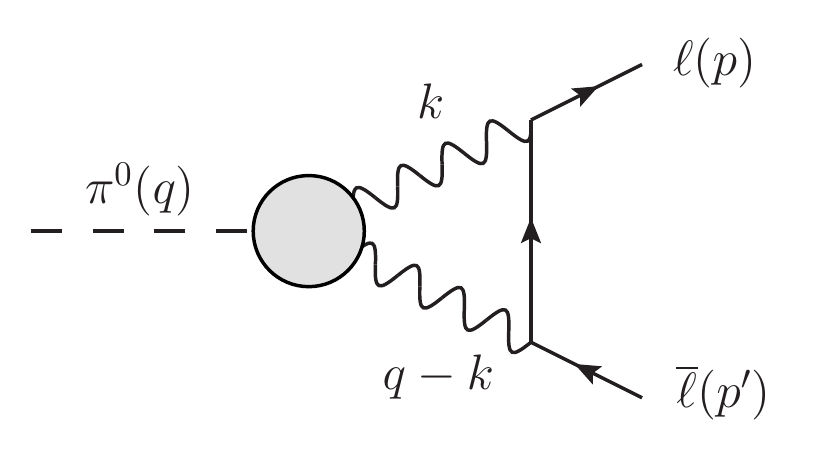}
\caption{\small{Feynman Diagram for $\pi^0 \to e^+e^-$ process.}}
\label{Pi0ee}
\end{figure}

Pseudoscalar decays into lepton pairs provide a unique environment for testing our knowledge of QCD. As such decays are driven by a loop process, they encode, at once, low and high energies. For the $\pi^0$ decay, the process (neglecting electroweak corrections) proceeds (Fig.~\ref{Pi0ee}) through the $\pi^0 \to \gamma^* \gamma^*$ anomalous vertex~\cite{Adler:1969gk}, with the photons linked by a lepton line. 
The loop does not diverge due to the presence of the $\pi^0$ transition form factor (TFF) on the anomalous vertex, the $F_{\pi^0\gamma^*\gamma^*}(k^2,(q-k)^2)$ with $k^2,(q-k)^2$ the photon virtualities. 
\vspace{-0.3cm}
\vspace{-0.3cm}
The TFF with $-q^2=Q^2$ cannot be calculated from first principles, and only the $Q^2=0$ and $Q^2\rightarrow\infty$ limits are known in terms of the axial anomaly in the chiral limit of QCD \cite{Adler:1969gk} and perturbative QCD \cite{Lepage:1980fj}, respectively. It has been customary to introduce a model for the TFF to interpolate between both regimes through the use of vector meson dominance ideas~\cite{Berman:1960zz,Bergstrom:1983ay,Dorokhov:2007bd} with different assumptions for describing the doubly virtual behavior~\cite{Dorokhov:2007bd}. The discrepancy among different choices there reflects the model-dependency of that procedure (see, for example, the discussion in Ref.~\cite{Dorokhov:2007bd}). In order to avoid this undesired model dependency, two main strategies have been developed so far based on the observed low-energy nature of the process (c.f. Eq.~\eqref{eq:loopaprox} and discussion afterwards). The first one involved the use of chiral perturbation theory ($\chi PT$) to calculate the diagram in Fig.~\ref{Pi0ee} (see~\cite{Knecht:1999gb} and references therein), where the low-energy constants of the theory would encode the high-energy effects from the process which are common, to first approximation, to $\pi^0$ and $\eta$ allowing to predict the $\pi^0$ decay in terms of the measured $\eta\rightarrow\mu^+\mu^-$ one~\cite{Agashe:2014kda}. The second one involved the use of a phenomenological parametrization of the TFF fitted to experimental data~\cite{Dorokhov:2007bd}. The lack of double virtual data for the double TFF however has been complemented so far with additional high-energy QCD constraints, which might distort the low-energy region fundamental in this process.

In this work, after briefly discussing the $\pi^0 \to e^+e^-$ process in Sec.~\ref{S2}, we explore for the first time the role of the bivariate Pad\'e approximants, the so-called Canterbury approximants (CA)~\cite{Chisholm} in Sec.~\ref{S3}, which will allow us to describe from the low-energies and in a model-independent data-driven approach the doubly virtual TFF driving this rare decay. CAs will allow us to evaluate the impact of the QCD high-energy tail in a consistent manner as well. We remark that the CA's technique can be generalized straightforwardly to other processes involving analytic Stieltjes and meromorphic functions with two variables. In the present work we exemplify its usage in a theoretically challenging process. Later on, in Sec.~\ref{S4} we will consider the impact of the KTeV measurement on the HLBL contribution to $(g-2)_{e,\mu}$ using the CAs. A potential new physics scenarios are considered in Sec.~\ref{S5} and the main results of our work are collected in the Conclusions.

\section{The $\pi^0 \to e^+e^-$ decay}~\label{S2}

The most accurate measurement of the $\pi^0 \to e^+e^-$ was performed by the KTeV Collaboration at Fermilab through the observation of almost 800 $\pi^0\to e^+e^-$ events~\cite{Abouzaid:2006kk} and yielded $\mathrm{BR}(\pi^0\to e^+e^-)=(7.48 \pm 0.29 \pm 0.25)\times 10^{-8}$ after removing the final state radiative corrections (RC)~\cite{Bergstrom:1982wk}, where the first error referred to statistics and the second to the total systematics.

The normalized $\mathrm{BR}(\pi^0\to e^+e^-)$ is defined as
\begin{equation}
\frac{\mathrm{BR}(\pi^0\rightarrow e^+e^-)}{\mathrm{BR}(\pi^0\rightarrow\gamma\gamma)}  =  2 \left(\frac{\alpha_{em} m_{e}}{\pi m_{\pi^0}}\right)^2\beta_{e} |\mathcal{A}(m_{\pi^0}^2)|^2,
\end{equation}
where $\beta_{e} = (1-4m_e^2/m_{\pi^0}^2)^{1/2}$ is the outgoing lepton velocity and $\mathcal{A}(m_{\pi^0}^2)$ is given by the loop integral
\begin{equation}
\mathcal{A}(q^2) =  2i \int \frac{\dd^4k}{\pi^2} \frac{ (q^2k^2 - (q k)^2)\tilde{F}_{\pi\gamma\gamma}(k^2,(q-k)^2)}{ q^2 k^2(q-k)^2((p-k)^2-m_e^2)} 
\label{eq:loop}
\end{equation}
and encodes all the hadronic effects through the normalized TFF $\tilde{F}_{\pi^0\gamma^*\gamma^*}(k^2,(q-k)^2)$ (i.e. $\tilde{F}_{\pi^0\gamma\gamma}(0,0)=1$).
Even without any information about the TFF, Cutcosky rules may be used to extract its imaginary part, which provides the well-known unitary bound discussed by Drell~\cite{Drell}, $\mathrm{BR}(\pi^0\to e^+e^-) \geq \mathrm{BR}^{\mathrm{unitary}}(\pi^0\to e^+e^-)=4.69\times 10^{-8}$, which is a model-independent result.

\begin{figure*}[hbt]
\includegraphics[width=0.45\textwidth]{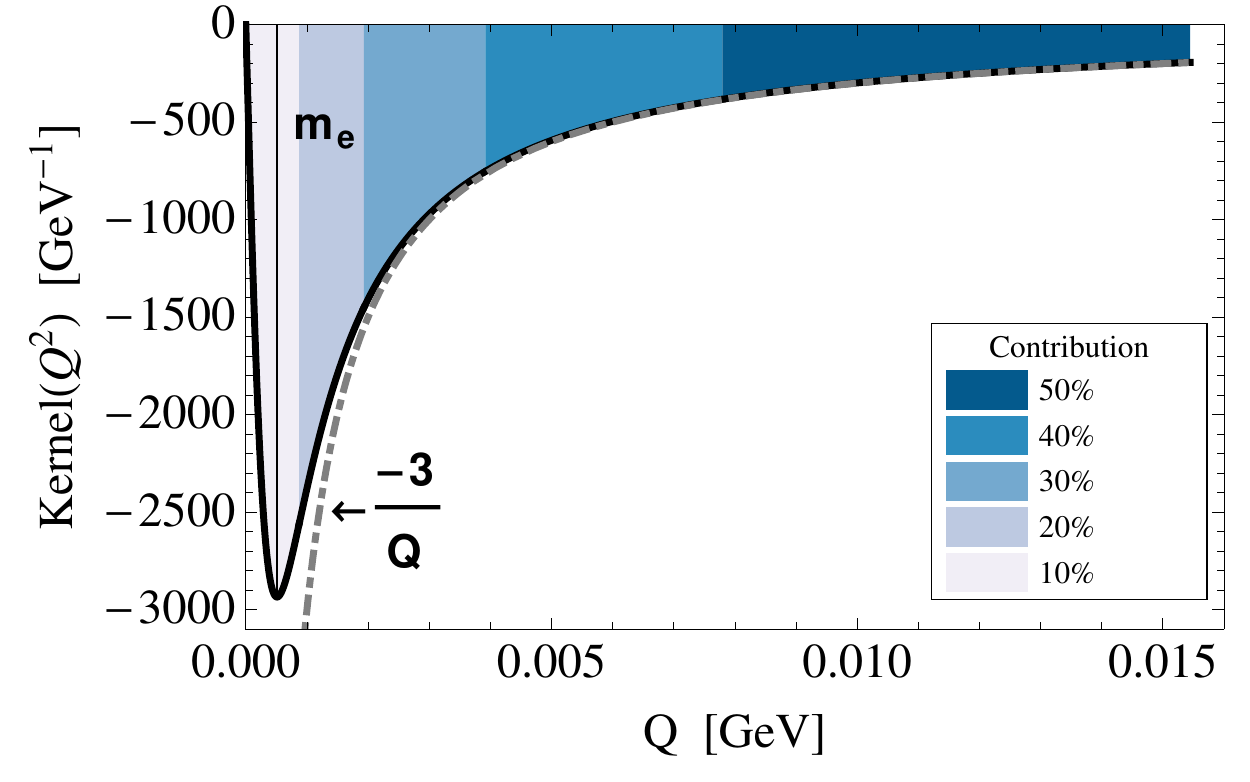}
\includegraphics[width=0.45\textwidth]{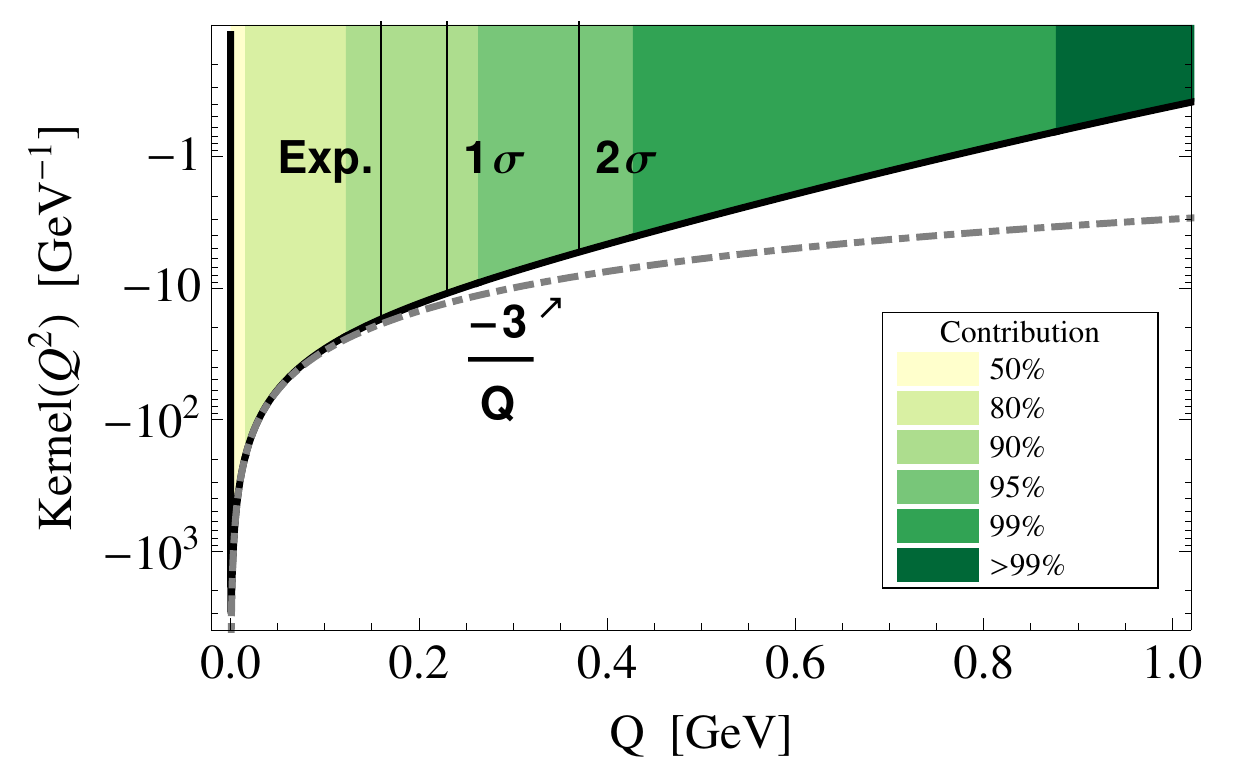}
\caption{\small{Kernel of the integral in Eq.~(\ref{eq:loopaprox}) for different energy ranges. Each band labels the partial contribution to the integral. }}
\label{figKernel}
\end{figure*}

The presence of the photon propagators (cf. Fig.~\ref{Pi0ee}) implies the kernel of the loop integral~(\ref{eq:loop}) to be peaked at very low energies of around the electron mass as is shown in Fig.~\ref{figKernel}. The kernel can be expanded in terms of $m_e/m_{\pi^0}$ as well as $m_e/\Lambda$ and  $m_{\pi^0}/\Lambda$, being $\Lambda$ the cut-off of the loop integral, or the hadronic scale driven by the TFF. 
Then, Eq.~(\ref{eq:loop}) reads~\cite{Knecht:1999gb,Dorokhov:2007bd}:
\begin{widetext}
\begin{equation}
\mathcal{A}(m_{\pi^0}^2) = \frac{i\pi}{2\beta_e}L + 
\frac{1}{\beta_{e}} \left( \frac{1}{4}L^2 
  +\frac{\pi^2}{12} + Li_2 \left(\frac{\beta_{e}-1}{1+\beta_{e}}\right)\right)
 -\frac{5}{4} + \int_0^{\infty}dQ\frac{3}{Q}\left(  \frac{m_e^2}{m_e^2+Q^2} -\tilde{F}_{\pi^0\gamma^*\gamma^*}(Q^2,Q^2) \right)
\label{eq:loopaprox}
\end{equation}
\end{widetext}
where $L= \ln\left(\frac{1-\beta_{e}}{1+\beta_{e}}\right)$ and terms of  $\mathcal{O} \left( \frac{m_e^2}{m_{\pi0}^2}, \frac{m_e^2}{m_{\pi^0}^2} \ln\frac{m_e^2}{m_{\pi^0}^2} \right)$ as well as $\mathcal{O}\left( \frac{m_e^2}{\Lambda^2} , \frac{m_e^2}{\Lambda^2} \ln \frac{m_{e}^2}{\Lambda^2}\right)$, and 
$\mathcal{O}\left( \frac{m_{\pi^0}^2}{\Lambda^2} , \frac{m_{\pi^0}^2}{\Lambda^2} \ln \frac{m_{e}^2}{\Lambda^2}\right)$ have been neglected.

The integral in Eq.~(\ref{eq:loopaprox}), Fig.~\ref{figKernel}, produces a negative contribution, diminishing the result. Omitting such contribution, 
Eq.~(\ref{eq:loopaprox}) would result in $19\times10^{-8}$ for the BR.

Recently, the authors of~\cite{Dorokhov:2007bd} resummed the power corrections using the Mellin-Barnes technique and found them of $\mathcal{O}(1\%)$~\cite{Dorokhov:2009xs}. Then, using a Vector Meson Dominance for the TFF, they found $\mathrm{BR}(\pi^0\to e^+e^-)=6.2(1)\times 10^{-8}$~\cite{Dorokhov:2007bd,Dorokhov:2009xs}, $3.2\sigma$ off the KTeV result.

Such discrepancy demands further explanations provided that future experiments (for example, the NA48/2 or NA62 kaon experiments at the CERN SPS which have demonstrated the possibility to perform precision $\pi^0$ decay physics~\cite{Goudzovski:2014rwa}) would confirm the current measurement. Three research lines can be conceived: a reevaluation of the radiative corrections, an improved parameterization of the doubly virtual TFF, or a new mechanism within physics beyond the SM~\cite{Kahn:2007ru,Chang:2008np}.

In Ref.~\cite{Vasko:2011pi}, the radiative corrections used by the KTeV based on Bergstr\"om's work~\cite{Bergstrom:1982wk} were reconsidered. At that time, Bergstr\"om considered the two-loop QED radiative correction to the decay in the soft-photon approximation together with an inclusion of a certain cut-off for the loop diagrams. He also considered the role of the Dalitz decay and its interference as a source of experimental error. The authors of~\cite{Vasko:2011pi} noticed that \cite{Bergstrom:1982wk} neglected a class of subleading diagrams, which due to particular cancelations among the dominant ones, turned out to be dominant. Later on, they also studied the role of the soft-photon approximation finding it accurate enough~\cite{Husek:2014tna}. Ref.~\cite{Bergstrom:1982wk} suggests that the radiative corrections represented a $-13\%$ effect, so increasing the value measured by KTeV (see~\cite{Abouzaid:2006kk} for details). The reanalysis of Refs.~\cite{Vasko:2011pi,Husek:2014tna} suggested, however, that including the subleading diagrams the RC would decrease down to $-6\%$, implying a smaller BR after RC are removed. With such considerations, the new KTeV value would result in $\mathrm{BR}_{\mathrm{"KTeV"}}(\pi^0\to e^+e^-)=(6.87 \pm 0.36)\times 10^{-8}$, closer to the SM value at about $2 \sigma$.

As we stated before, we investigate here the role of the TFF on such decay. Given the accuracy of Eq.~(\ref{eq:loopaprox}), it is safe to conclude that the main contribution to the loop integral~(\ref{eq:loop}) happens at low space-like energies (Fig.~\ref{figKernel}). Therefore, a precise description of the doubly virtual TFF at low space-like energies is an essential starting-point for an accurate prediction, an observation overlooked so far. Moreover, we noticed that the factorization approximation for the TFF, i.e., $F(Q_1^2,Q_2^2)=F(Q_1^2,0)\times F(0,Q_2^2)$, can induce large effects. 
For our study, we consider the reconstruction of the TFF of double virtuality by a data driven approach. That method, model independent, is based on the theory of Pad\'e approximants (PA)~\cite{Baker} extended to the double virtual case, the Canterbury approximants~\cite{Chisholm}. 

\section{Canterbury approximants}\label{S3}

Canterbury approximants (CA) are rational approximants defined from double power series in variables $Q_1^2$ and $Q_2^2$ with the following properties~\cite{Chisholm}: 
\begin{enumerate}
\item CA are symmetric with respect to $Q_1^2$ and $Q_2^2$.
\item CA reduce to the PA if $Q_1^2(Q_2^2)=0$.
\item CA enjoy homographic invariance under argument transformations. 
\item CA are unique.
\item CA satisfy the low-energy constraints by definition.
\item CA easily accommodate the high-energy QCD constraints.
\end{enumerate}

Even though generalizations of CA are known~\cite{Levin,HughesJones,Cuyt,Guillaume}, they will reduce to CA for the $\pi^0\to e^+e^-$, representing the simplest and most efficient model-independent approach. On top, the convergence to symmetric two-variable Stieltjes functions is known to exist~\cite{Chisholm,Alabiso:1974vk}. In our case of study, the TFF, the first property follows trivially from Bose symmetry.
On top, when one of the photon virtualities is zero, the resulting TFF can be safely approximated by a standard PA with great success~\cite{Masjuan:2012wy,Masjuan:2012qn,Escribano:2013kba}, satisfying the second CA property. The third property is used in~\cite{Chisholm} together with the accuracy-through-order conditions to explore the convergence of the CA. As it is shown below, properties 4), 5) and 6) emerge naturally by construction. The important remark here is that the TFF of doubly virtuality \cite{Masjuan:2012wy,Masjuan:2012qn,Escribano:2013kba} is a bivariate Stieltjes function~\cite{Alabiso:1974vk} for which the CA's convergence theorem from~\cite{Chisholm,Alabiso:1974vk} apply.

CA, defined as $C^N_N(x,y) =  \frac{R_N(x,y)}{Q_N(x,y)}$~\cite{Chisholm}, extend PAs from one to two variables. The coefficients of $R_N(x,y)=\sum_{i,j}^{N}  a_{i,j}x^iy^j$, $(i,j) \in {\cal N}$, and $Q_N(x,y)=\sum_{i,j}^{N}  b_{i,j}x^iy^j$, $(i,j) \in {\cal D}$ are determined by the accuracy-through-order conditions up to order  ${\cal O}(x^{2N}y^{2N} )$, and ${\cal O}(x^{2N+1-\alpha}y^{\alpha} )$ with $\alpha=0,1,\dots 2N$, i.e., given $f(x,y)=\sum_{i,j} c_{i,j}x^iy^j$,
\begin{equation} \nonumber
\sum_{i=0}^{\alpha}\sum_{j=0}^{\beta}b_{i,j} c_{\alpha-i,\beta-j} = a_{\alpha,\beta} \quad \textrm{ for } (\alpha,\beta) \in {\cal N}
\end{equation}
and ($b_{0,0}=1$ as part of the definition)
\begin{equation}\nonumber
\sum_{i=0}^{\textrm{min}(\alpha,N)}\sum_{j=0}^{\textrm{min}(\beta,N)}b_{i,j} c_{\alpha-i,\beta-j} =0 \, \textrm{for } (\alpha,\beta) \in {\cal E}, (\alpha,\beta) \notin {\cal N}
\end{equation}
\noindent
where $\textrm{dim}({\cal E})=\textrm{dim}({\cal N})+\textrm{dim}({\cal D})-1$. With $x=Q_1^2$ and $y=Q_2^2$, we guarantee, as anticipated, a correct low-energy description. The Canterbury group~\cite{Chisholm,Levin,HughesJones} demonstrated that for Stieltjes and meromorphic bivariate functions, the convergence of $C^{N+J}_N(Q_1^2,Q_2^2)$ is guaranteed for $J=-1,0$, properties exploited here~\cite{prep}.

The first element on the CA sequence reads
\begin{equation}\label{Chisholm}
C^0_1(Q_1^2,Q_2^2)=\frac{a_{0,0}}{1+b_{1,0}(Q_1^2+Q_2^2)+b_{1,1}Q_1^2Q_2^2}\, ,
\end{equation}
\noindent
where Bose symmetry is already implemented ($b_{i,j}=b_{j,i}$) and $b_{0,0}=1$ without loss of generality. Eq.~(\ref{Chisholm}) reproduces the high-energy behavior when one photon virtuality is set to zero, the well known Brodsky-Lepage limit~\cite{Lepage:1980fj}, property $5$) above. Knowing the Taylor expansion of the $F(Q_1^2,Q_2^2)$, Eq.~(\ref{Chisholm}) would be unique: $a_{0,0}=F(0,0)$ is determined from the $\Gamma (\pi^0 \to \gamma\gamma)$ through the relation $(4\pi \alpha)^2 m_{\pi}^3 a_{0,0}^2=64 \pi \Gamma (\pi^0 \to \gamma\gamma)$; $b_{1,0}$ is the slope of the single virtual $\pi^0$-TFF $b_{\pi}$; and $b_{1,1}$ is related to the doubly-virtual slope.  

The second element on the CA sequence with the appropriate high-energy behavior results in
\begin{widetext}
\begin{equation}\label{Chisholm2}
C^1_2(Q_1^2,Q_2^2)=\frac{a_{0,0} +a_{1,0}(Q_1^2+Q_2^2)+a_{1,1}Q_1^2Q_2^2}{1+b_{1,0}(Q_1^2+Q_2^2)+b_{1,1}Q_1^2Q_2^2+b_{2,0}(Q_1^4+Q_2^4)+b_{2,1}(Q_1^4 Q_2^2+Q_1^2 Q_2^4)  +b_{2,2} Q_1^4 Q_2^4}\, ,
\end{equation}
\end{widetext}
and demands the knowledge of four coefficients belonging to the double virtual sector ($a_{1,1}, \ b_{1,1}, \ b_{2,1}$ and $b_{2,2}$) as well as the curvature and third derivative of the single virtual TFF to match $a_{1,0}$ and $b_{2,0}$. Obeying the convergence properties of the CA, $C^1_2(Q_1^2,Q_2^2)$ approximates better the TFF than $C^0_1(Q_1^2,Q_2^2)$. The difference among them can be taken as a way to estimate the systematic error on the approximation sequence~\cite{prep}.


Experimental data for the doubly virtual $F(Q_1^2,Q_2^2)$ are not available yet and we cannot extract all those terms from them. To yield a result with our method we need to find a compromise. On the one hand, the OPE tells us that $\lim_{Q^2 \to \infty} F(Q^2,Q^2) \sim Q^{-2}$ and implies $b_{1,1}=0$ in Eq.(\ref{Chisholm}) and $b_{2,2}=0$ in Eq.(\ref{Chisholm2}). If the OPE is fulfilled, no experimental information about the doubly virtual TFF is required up to the second element on the sequence, the $C^1_2(Q_1^2,Q_2^2)$. On the other hand, if we impose the factorization approach to $C_1^0(Q_1^2,Q_2^2)$, which means $C_1^0(Q_1^2,Q_2^2) = P_1^0(Q_1^2) \times P_1^0(Q_2^2)$, we would find $b_{1,1}= b_{1,0}^2$. In this second scenario, the high-energy tail of the TFF would behave as $Q^{-4}$ instead. However, the $\chi$PT study performed in Ref.~\cite{Bijnens:2012hf} favors the factorization approach at very-low energies for the TFF, indicating that the role of the high-energy tail may not be that relevant at the low scales which are crucial in our calculation~\cite{Husek:2014tna}. In summary, in order to remain on the conservative side, we assume for the $C^0_1(Q_1^2,Q_2^2)$ approximant the ---yet unknown--- $b_{1,1}$ value to lie within the reasonable $0 \leq b_{1,1} \leq b_{1,0}^2$ range, which represents a compromise between the sought low-energy description and the high-energy QCD constraint. The lower limit coincides with the OPE constraint and prevents as well the $C_1^0(Q_1^2,Q_2^2)$ to have a divergency on the space-like region; the upper one is the value implied by factorization. In calculating the BR we do not use~(\ref{eq:loopaprox}), but~(\ref{eq:loop}), including in addition the SM $Z$-boson contribution (of order $-0.02 \times 10^{-8}$). With such parameters, and taking the $b_{\pi}$ value obtained in~\cite{Masjuan:2012wy}, we obtain $\textrm{BR}_{\mathrm{SM}}(\pi^0\to e^+e^-)=(6.20 - 6.35) (4)\times 10^{-8}$ where the two main numbers come from ranging $b_{1,1}$, and the error comes from $\Gamma (\pi^0 \to \gamma\gamma) $ and $b_{\pi}$ uncertainties.

The next element of the sequence, the $C^1_2(Q_1^2,Q_2^2)$, demands (c.f Eq.~(\ref{Chisholm2})) not only imposing the $\lim_{Q^2 \to \infty} F(Q^2,Q^2) \sim Q^{-2}$ but also its exact coefficient. 
For that reason, to extract $a_{1,1}$ and $b_{2,1}$ in ~(\ref{Chisholm2}), we take $\lim_{Q^2 \to \infty} F(Q^2,Q^2) = \frac{2F_{\pi}}{3Q^{2}}\left( 1 - \frac{8\delta^2}{9Q^2} +{\cal O}(Q^{-4})\right) $ where $\delta^2 =0.20(2)$GeV$^{2}$ from Ref.~\cite{Novikov:1983jt}. The remaining parameter in~(\ref{Chisholm2}), the $b_{1,1}$, is associated to the low-energy expansion 
$F_{P\gamma^*\gamma^*}=F_{P\gamma\gamma}(1 + \frac{b_{\pi}}{m_{\pi}^2}(Q_1^2+Q_2^2) + ... + \frac{a_{\pi;1,1}}{m_{\pi}^4}Q_1^2Q_2^2 + ... )$, i.e., with the yet unknown slope of doubly virtuallity $a_{\pi;1,1}$.
To obtain $b_{1,1}$, we determine $a_{\pi;1,1}$ similarly as we did for the $C^0_1(Q_1^2,Q_2^2)$ case before which is, after matching all the other parameters we let this one range freely with the only constraint of not developing divergencies in the space-like region. In this respect, we find $1.92 b_{\pi}^2 \leq a_{\pi;1,1} \leq 2.07 b_{\pi}^2$. Experimental data will ultimately decide on this issue since knowing both the low-energy parameters and the OPE we will reach higher order in the $C^N_{N+1}(Q_1^2,Q_2^2)$ sequence and the discussion about OPE versus factorization will be unnecessary. Notice that the $C^1_2(Q_1^2,Q_2^2)$, even though has its parameters fixed by OPE constraints, still fulfills the factorization approach at low energies as discussed in Ref.~\cite{Bijnens:2012hf}. 

With all these constraints, we determine a new SM range 
\begin{equation}
\textrm{BR}_{\mathrm{SM}}(\pi^0\to e^+e^-)=(6.22 - 6.23) (4)(2)\times 10^{-8}\, ,
\label{eq:SMresult}
\end{equation}
\noindent
where the two main numbers come from the ranging of $a_{\pi;1,1}$. As before, the error $\pm 4 \cdot 10^{-8}$ comes from $\Gamma (\pi^0 \to \gamma\gamma) $ and $b_{\pi}$ uncertainties, and the $\pm 2\times 10^{-8}$ from the evaluation of the systematic error of our approximation. Such error is obtained after comparing the difference between the $C^0_1(Q_1^2,Q_2^2)$ and $C^1_2(Q_1^2,Q_2^2)$~\cite{Masjuan:2012wy,prep}. The result in Eq.~(\ref{eq:SMresult}) can be summarized as BR$_{\mathrm{SM}}(\pi^0\to e^+e^-)=6.23(4)(3)\times 10^{-8}$ with the systematic and the spread in Eq.~(\ref{eq:SMresult}) combined linearly. This results reduces the error on the previous SM determination by more than $50\%$, which demands now including the $Z$ boson contribution and makes no approximation on the calculation of the loop integral. Given that our final error is dominated by the input errors and not by the observed convergence of our approximants, we anticipate that the next element, the $C^2_3(Q_1^2,Q_2^2)$, would not provide any valuable information. In this respect, we believe that the expected experimental measurement of the slope of doubly virtuality will not make a dramatic change with respect to Eq.~(\ref{eq:SMresult}) beyond reducing the range quoted (unless we also find a surprise there).

For completeness, let us indicate that if we would have considered approximation~\eqref{eq:loopaprox} instead of the full result derived from~\eqref{eq:loop}, we would have obtained 
$(6.17-6.18)\times10^{-8}$, while omitting the $Z$ boson contribution would
lead to $(6.24-6.25)\times10^{-8}$. Eq.~\eqref{eq:SMresult} still represents a deviation of the measured BR of about $3.2 \sigma$ (or $1.8\sigma$ with RC in~\cite{Vasko:2011pi,Husek:2014tna}), which we think is unsurmountable in a model-independent way with our present knowledge of the Standard Model. Therefore, to eventually improve on the situation, experimental data would be, as said, required.

\section{Impact of the KTeV measurement on the HLBL contribution to $(g-2)_{e,\mu}$}\label{S4}

CAs are flexible enough to be able to reproduce, for the first time, the KTeV measurement ---which translates into tuning  $a_{1,1}, b_{1,1}$ in Eq.(\ref{Chisholm2})--- if our previous scenarios (i.e., factorization and high-energy QCD constraints) are momentarily put aside, a procedure which does not spoil any property from Pad\'e Theory \cite{Masjuan:2007ay}. 

\begin{figure*}[htbp]
\begin{center}
\includegraphics[width=8.5cm]{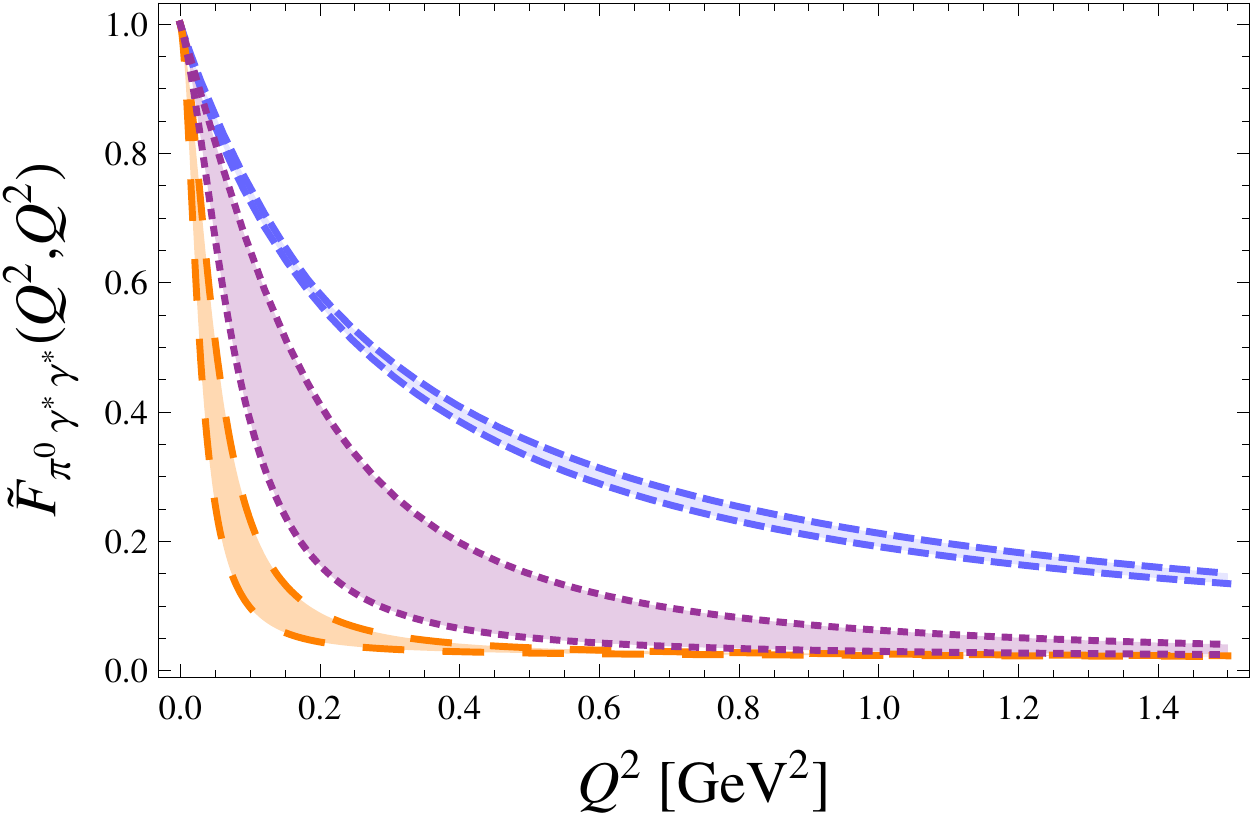}
\includegraphics[width=8.5cm]{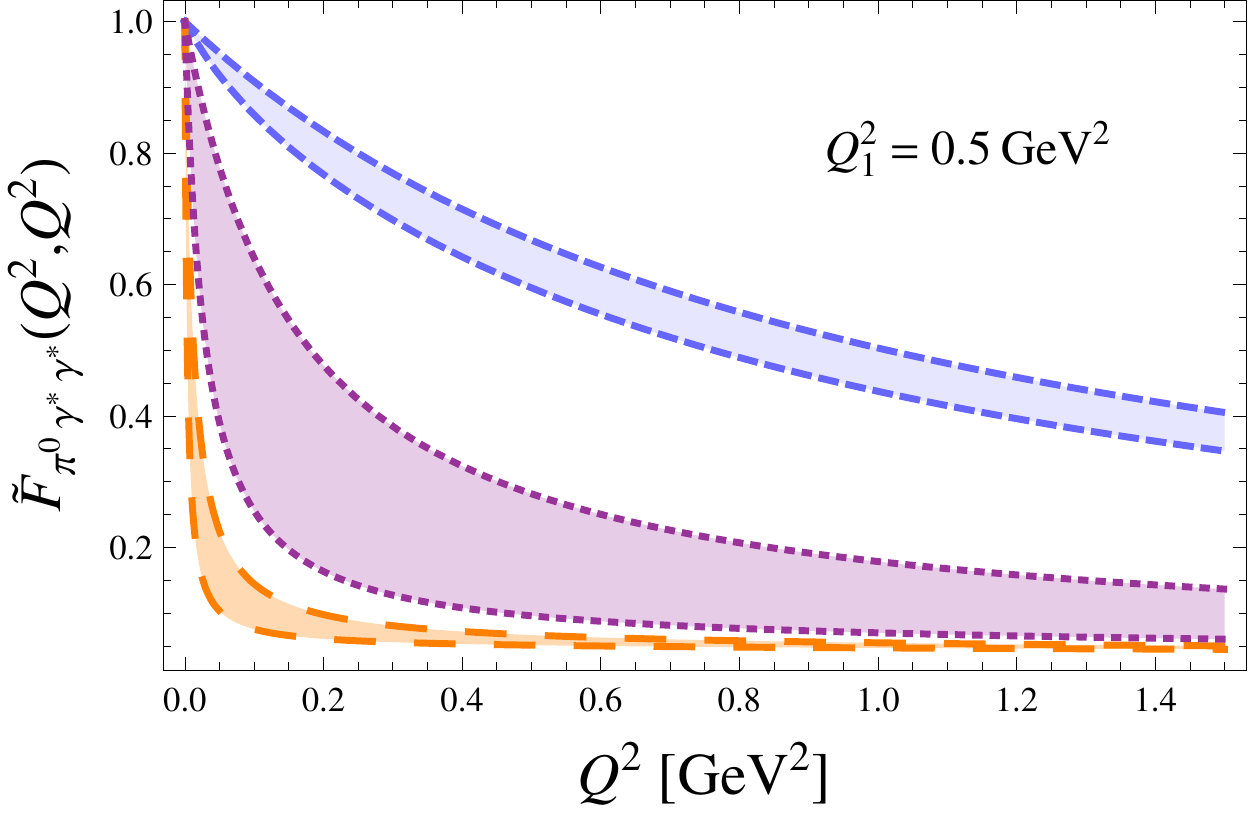}
\caption{\small{Left panel: normalized TFF assuming $Q_1^2 = Q_2^2=Q^2$. Right panel: normalized TFF assuming $Q_1^2=0.5$ GeV$^2$. Black solid line indicates the factorized TFF. Upper (blue) band shows our $C^1_2(Q_1^2,Q_2^2)$ estimation with $1.92b_{\pi}^2 \leq a_{\pi;1,1} \leq 2.07 b_{\pi}^2$. Lower (orange) band reproduces the KTeV measurement within 1$\sigma$. Middle (purple) band considers KTeV measurement with the new RC }}
\label{figTFF}
\end{center}
\end{figure*}

This result is shown in Fig.~\ref{figTFF} as a purple band when including the new RC,  for which $-39 b_{\pi}^2 \leq a_{\pi;1,1} \leq-4 b_{\pi}^2$. For completeness, we show as well what would have been achieved without the new RC as an orange band (which corresponds to $-370 b_{\pi}^2 \leq a_{\pi;1,1} \leq -100 b_{\pi}^2$). While the very low-energy region is not very different from our result \eqref{eq:SMresult} (blue band), KTeV seems to imply a strongly decreasing TFF at large $Q^2$. This effect could be achieved with a very slow converging OPE as indicated by the large $\delta^2\gtrsim 10~\textrm{GeV}^2$ value we obtained through matching $a_{1,1}$ and $b_{1,1}$ to KTeV. This feature has never been observed before.

A first glance on Fig.~\ref{figTFF} reveals that, even without high precision data at a given $Q^2$ ($30\%$ or even $50\%$ experimental errors), these 3 scenarios could be easily distinguished, while improving our result~(\ref{eq:SMresult}) would require higher precision, around $(10-20)\%$. Such data may be fitted through CAs, which being able to accommodate the high-energy constraints from QCD as well, would allow to reconstruct the TFF from $Q^2=0$ to $\infty$ in the space-like region.

Given that nowadays the KTeV measurement represents then the only (indirect) source of experimental information on $b_{1,1}$, 
as an amusement, we can use the results from the previous paragraph and explore the impact of KTeV on the hadronic Light-by-Light contribution to $(g-2)_{e,\mu}$. There, the $\pi^0$ yields the leading contribution, $a_{e,\mu}^{HLBL;\pi^0}$, and thereby the $\pi^0$-TFF plays a central role in the calculations~\cite{Jegerlehner:2009ry}. Taking the range given by the purple band, we find for the (off-shell) $\pi^0$-exchange $a_{e}^{\textrm{HLBL};\pi^0}=2.0(3) \times 10^{-14}$, where the error accounts for the band's width, 
(to compare with $3.0(3)\times10^{-14}$~\cite{Jegerlehner:2009ry}). For the (on-shell) $\pi^0$-pole we obtain $a_{e}^{\textrm{HLBL};\pi^0}=1.8(2) \times 10^{-14}$ (to compare with $2.6\times10^{-14}$~\cite{Knecht:2001qf}).

Given that $a_e^{\textrm{exp}} - a_e^{\textrm{th}} = -105(81)\times10^{-14}$~\cite{Jegerlehner:2009ry,PDG2014}, this effect is negligible though in the correct direction. This contrasts with its muon counterpart, which in addition is more sensitive to hadronic physics. There, the deviation $a_{\mu}^{\textrm{exp}} - a_{\mu}^{\textrm{th}} = +290(90)\times10^{-11}$~\cite{Jegerlehner:2009ry,PDG2014} has opposite sign. For the (off-shell) $\pi^0$-exchange we obtain $a_{\mu}^{\textrm{HLBL};\pi^0}=40(9) \times 10^{-11}$ (to compare with $72(12)\times10^{-11}$~\cite{Jegerlehner:2009ry}). For the (on-shell) $\pi^0$-pole we obtain $a_{\mu}^{\textrm{HLBL};\pi^0}=36(7) \times 10^{-11}$ (to compare with $58(10)\times10^{-11}$~\cite{Knecht:2001qf}). These shifts represents twice the foreseen experimental accuracy $(16\times10^{-11})$ in the future $(g-2)_{\mu}$ experiments projected in Brookhaven and J-PARC ~\cite{Roberts:2010cj,Iinuma:2011zz} indicating that the current precision of the SM error on the $(g-2)_{\mu}$ would be underestimated if the $\pi^0\rightarrow e^+e^-$ would be taken into account. Taking KTeV results without the latest RC would accentuate the differences indicated above. A thorough application of our approach to $(g-2)$ will be discussed elsewhere.

\section{New physics contributions to $\pi^0 \to e^+e^-$}\label{S5}

As anticipated in the introduction, tree level contributions from new physics may be relevant for this process as well. In our case, only pseudoscalar $(P)$ and axial ($A$) channels are relevant, since interactions such as leptoquarks may be Fierz-rearranged~\cite{Soni:1974aw}, then, only pseudoscalar and axial channels will appear. The following Lagrangian
\begin{equation}
 \label{eq:NPL}
\mathcal{L} = \frac{-g}{4m_W} \sum_{f} m_Ac^{A}_{f} \left(\overline{f}\slashed{A}\gamma_5f \right) + 2 m_f c^{P}_{f} \left(\overline{f}i\gamma_5 f \right)P, 
\end{equation}
where $g, m_W$ are the standard electroweak parameters, and $c^{A(P)}_{f}$ are dimensionless couplings to the fermion $f=\{u,d,e\}$, yields an additional term modifying Eq.~(\ref{eq:loop})
\begin{equation}
   \label{eq:NP}
   \mathcal{A}(q^2) \rightarrow  \mathcal{A}(q^2) + \frac{\sqrt{2} G_F F_{\pi}}{4\alpha_{em}^2 F_{\pi^0\gamma\gamma}(0,0)}(\lambda^A+\lambda^P) , 
\end{equation}
where $G_F$ is the Fermi coupling constant, $F_{\pi}$ the pion decay constant, $\lambda^A = c^A_e \left( c^A_u - c^A_d \right)$ and  $\lambda^P = c^P_e \left( c^P_u - c^P_d \right)/(1-m_P^2/m_{\pi^0}^2)$. As an illustration, the $Z^0$ boson ($c^A_{e,d} = - c^A_u = -1,c^P_f=0$) shifts $\mathcal{A}(q^2)$ by $-0.3\%$. Similarly, the ``dark'' $Z$ model in~\cite{Davoudiasl:2012qa}, would reduce the amplitude by $(-0.3\%) \delta^2$, with $\delta^2 \ll 1$~\cite{Davoudiasl:2012ag}. Therefore, electroweak-like effects are unlikely to be important and more general NP approaches would be required when considering axial interactions~\cite{Kahn:2007ru}. Besides, pseudoscalar contributions appear in extended Higgs sectors, such as the supersymmetric model in Ref.~\cite{Chang:2008np}. Their effects in our decay may become large enough if $m_{P} \simeq m_{\pi^0}$ as may be inferred from the pole in $\lambda^P$, though existing constraints on $m_P$ discard this scenario. Since many phenomenological constraints to these simple NP scenarios exist~\cite{Davoudiasl:2012ag} (dark photon searches, $(g-2)_{e,\mu}$, electroweak physics), a suitable scenario represents a challenging study which is beyond the scope of this article.

\section{Conclusions}\label{Conclusions}

In this work, we presented a model-independent approach to describe symmetric bivariate functions based on Canterbury approximants. This method allows a model-independent study of the doubly virtual $\pi^0$-TFF in the space-like region which incorporates the low- and high-energy QCD constraints. These are essential prerequisites for calculations such as $\pi^0\rightarrow e^+ e^-$ or $a_{\mu}^{HLBL;\pi^0}$. We have predicted BR$_{\mathrm{SM}}(\pi^0\to e^+e^-)=6.23 (4)(3)\times 10^{-8}$, still $2\sigma$ off the experimental result and found that, unless New Physics are present, the KTeV result on $\pi^0\rightarrow e^+ e^-$ implies an unexpected behavior for the doubly virtual TFF. This would produce a large shift for $a_{\mu}^{HLBL;\pi^0}$, of the order of projected experiments for measuring $(g-2)_{\mu}$. The effect of CAs in $\pi^0\rightarrow 4\ell$ decays is studied in~\cite{Adlarson:2014hka}. Possible effects may appear as well in $e p(e n)$ elastic scattering, which would be relevant for the proton radius puzzle. 

We conclude that the current situation demands an experimental measurement of $F_{\pi^0\gamma^*\gamma^*}(Q_1^2,Q_2^2)$ as well as a new $\pi^0\rightarrow e^+e^-$ determination.

\section*{Acknowledgements}

We would like to thank T. Beranek, K. Kampf, A. Kupsc, S. Leupold, M. Knecht, and M. Vanderhaeghen for encouragement and discussions. Work supported by the Deutsche Forschungsgemeinschaft DFG through the Collaborative Research Center ``The Low-Energy Frontier of the Standard Model" (SFB 1044).

\end{document}